\documentclass[a4paper, amsfonts, amssymb, amsmath, reprint, showkeys,
twoside, aps, superscriptaddress]{revtex4-2}
\usepackage[margin=1in]{geometry} 
\usepackage{xcolor}
\usepackage{graphicx}
\usepackage[version=4]{mhchem}
\usepackage[pdftex, pdftitle={Article}, pdfauthor={Author}]{hyperref}
\usepackage{listings}
\usepackage{physics}
\usepackage[T1]{fontenc}
\usepackage[english]{babel}
\usepackage{bookman}
\usepackage{footmisc}

\usepackage{soul} 

\newcommand{\FIG}[1]{#1}

\usepackage[textwidth=0.75in]{todonotes}
\DeclareMathOperator*{\argmin}{argmin}
\newcommand{\DC}{DC}
\newcommand{\TDC}{TDC}

\begin{document}
\title{\scshape An unambiguous and robust formulation for Wannier localization}

\author{Kangbo Li}
\affiliation{Department of Computer Science, Cornell University, Ithaca, NY 14853, USA}
\author{Hsin-Yu Ko}
\affiliation{Department of Chemistry and Chemical Biology, Cornell University, Ithaca, NY 14853, USA}
\author{Robert A. DiStasio Jr.}
\email[Corresponding author: ]{distasio@cornell.edu}
\affiliation{Department of Chemistry and Chemical Biology, Cornell University, Ithaca, NY 14853, USA}
\author{Anil Damle}
\email[Corresponding author: ]{damle@cornell.edu}
\affiliation{Department of Computer Science, Cornell University, Ithaca, NY 14853, USA}


\date{\normalsize \today}

\begin{abstract} 
%
%
%
\noindent We provide a new variational definition for the spread of an orbital under periodic boundary conditions (PBCs) that is continuous with respect to the gauge, consistent in the thermodynamic limit, well-suited to diffuse orbitals, and systematically adaptable to schemes computing localized Wannier functions. Existing definitions do not satisfy all these desiderata, partly because they depend on an “orbital center”—an ill-defined concept under PBCs. Based on this theoretical development, we showcase a robust and efficient (10x-70x fewer iterations) localization scheme across a range of materials.
\end{abstract}

\maketitle

Localized Wannier functions (LWFs) offer several theoretical and computational advantages over canonical (Bloch) orbitals in electronic structure calculations of condensed-phase systems.
For one, LWFs allow one to probe/characterize the local electronic structure in complex extended systems, thereby enabling chemical bonding analysis (e.g., floating bonds in amorphous \ce{Si}~\cite{fornari2001wannier,MaximallyLocalSilves1998} and \ce{Si}-\ce{C} alloys~\cite{fitzhenry2002wannier}), orbital partitioning (e.g., for computing molecular dipole moments in liquid water~\cite{silvestrelli1999water}), and the construction of chemical-environment-based features and/or targets for machine learning~\cite{zhang2020deep}. 
LWFs also play a critical role in evaluating the bulk properties of materials (e.g., in the modern theory of polarization~\cite{TheoryOfTheEResta1992, MacroscopicPolResta1994, TheoryOfPolarKingS1993, ElectricPolariVander1993} and magnetization~\cite{OrbitalMagnetiThonha2005, BerryPhaseCorXiao2005, OrbitalMagnetiCereso2006, QuantumTheoryShiJ2007,DichroicMmlMaSouza2008}), predicting and understanding the spectroscopic signatures of condensed matter (e.g., IR~\cite{IntermolecularSharma2005} and Raman spectra~\cite{RamanSpectraOWanQ2013}), as well as constructing effective model Hamiltonians (e.g., for quantum transport of electrons~\cite{IAbInitioICalzol2004, BandStructureLeeY2005} and strongly correlated systems~\cite{TamingMultipleFabris2005, OrbitalDensityAnisim2007, ScreenedCoulomMiyake2008, OptimalRepreseUmari2009}).
Computationally, LWFs also enable large-scale electronic structure calculations to exploit the inherent real-space sparsity (or ``near-sightedness''~\cite{prodan_nearsightedness_2005}) of exchange-correlation interactions (e.g., in hybrid DFT~\cite{EnablingLargeKoHs2020,EnablingLargeKoHs2021} and GW~\cite{IAbInitioIButh2005}).
As such, a well-defined, robust, and computationally efficient theoretical and algorithmic framework for obtaining LWFs remains highly desirable.

In this Letter, we address two fundamental problems with existing methodologies for obtaining LWFs---one theoretical and one practical.
%
First, we present a new variational definition for the spread of an orbital under PBCs that has several favorable properties, including: continuity with respect to gauge transforms, consistency in the thermodynamic limit, and applicability to diffuse orbitals.
%
Most importantly, our formulation provides an \emph{unambiguous} ground-truth definition of orbital spread; it sidesteps the fundamentally ill-posed problem of defining an orbital center under PBCs and doing so allows the definition to be unambiguous and gauge continuous.
In contrast, prior work based on adaptations of $\langle r^2\rangle - \langle r \rangle^2$ (e.g., Marzari-Vanderbilt~\cite{MaximallyLocalMarzar1997}) yield orbital spread ans\"{a}tze that are not gauge continuous (a property closely related to the complexity of defining a position operator under PBCs as studied by Resta~\cite{QuantumMechaniResta1998,ElectricalPolaResta2010}).
Given the importance of LWFs, many practically effective ``definitions'' (or ``approximations'') for an orbital spread under PBCs have been proposed~\cite{QuantumMechaniResta1998, GeneralAndEffBergho2000, PartlyOccupiedThyges2005, MaximallyLocalSilves1998, MaximallyLocalSilves1999, AccuratePolariStenge2006}. 
While many of these proposals were deemed ``equivalent''~\cite{MaximallyLocalMarzar2012, SpreadBalancedFontan2021} for generating LWFs, we will show this is not the case.



In practice, our definition of orbital spread could be used directly within a localization scheme.
%
However, it is inherently ``global'' in $k$-space and that would make such a scheme marginally more expensive than prior work per iteration. 
Therefore, we develop a systematic approximation to our spread formulation that is local in $k$-space and derive its gradient. 
When paired with techniques from manifold optimization, our approximation results in a favorable localization scheme because: (1) it can be systematically improved if desired, (2) it has an explicit relation to an unambiguous ground truth formula and, therefore, the approximation errors can be quantified, (3) it retains the continuity with respect to gauge transforms of our variational definition, (4) it naturally spans both $k$-point sampling and $\Gamma$-point only calculations, and (5) it exhibits superior performance to prior work. 
%
In particular, we show that our method consistently converges in at least an order-of-magnitude fewer iterations ($10\times$$-$$70\times$) than the widely used \texttt{Wannier90} code~\cite{Wannier90AsAPizzi2020} across a diverse set of materials.

Our definition for the spread of an orbital under PBCs is motivated by the variational characterization of the center of an orbital under \textit{open} boundary conditions:
\begin{align}
\mathbf{c}^* &\triangleq \argmin_{\mathbf{r}'} 
\int \rho(\mathbf{r}) (\mathbf{r} - \mathbf{r}')^2 \mathrm{d} \mathbf{r} = \int \rho(\mathbf{r}) \mathbf{r} \, \mathrm{d} \mathbf{r} \label{eqn:variational_center} ,
\end{align}
in which $\rho(\mathbf{r})$ is the density of a given Wannier function.
Adapting~\eqref{eqn:variational_center} for PBCs leads us to the following \emph{density convolution} (\DC), denoted by $\mathcal{S} \rho$:
\begin{align}
(\mathcal{S} \rho) (\mathbf{r}') &\triangleq 
\int_{\mathbb{S}_{\mathbf{r}'}} \rho(\mathbf{r}) (\mathbf{r} - \mathbf{r}')^2 \mathrm{d} \mathbf{r} \nonumber \\
&= \int_{\mathbb{S}_{\mathbf{0}}} \rho(\mathbf{r} + \mathbf{r}') \mathbf{r}^2 \mathrm{d} \mathbf{r},
\label{eqn:density_convolution}
\end{align}
where $\mathbb{S}_{\mathbf{0}}$ is the Born–von Karman supercell (i.e., the unit cell replicated with respect to the $\mathbf{k}$-point mesh) and $\mathbb{S}_{\mathbf{r}'}$ is the supercell translated by $\mathbf{r}'$.
The \DC~center ($\mathbf{c}_{\textsc{dc}}$) and \DC~spread ($s_{\textsc{dc}}$) are then defined as the minimizer and minimum of~\eqref{eqn:density_convolution} respectively:
\begin{align}
    \mathbf{c}_{\textsc{dc}} &\triangleq \argmin_{\mathbf{r}'} (\mathcal{S} \rho) (\mathbf{r}') \label{eqn:density_convolution_center} \\
    s_{\textsc{dc}} &\triangleq \min_{\mathbf{r}'} (\mathcal{S} \rho) (\mathbf{r}'). \label{eqn:density_convolution_spread}
\end{align}

A key feature of $s_{\textsc{dc}}$ is that it is continuous with respect to the gauge chosen for the Wannier functions.
This property arises because $s_{\textsc{dc}}$ is explicitly defined as the minimum of an optimization problem---a continuous quantity with respect to the gauge in this case.
This is in stark contrast to commonly used definitions for orbital spread based on $\langle r^2\rangle - \langle r \rangle^2$ (e.g.,~\cite{MaximallyLocalMarzar1997}) that can be \emph{discontinuous} with respect to the gauge---a property inherited from explicitly depending on the ill-defined notion of an orbital center under PBCs~\cite{QuantumMechaniResta1998,ElectricalPolaResta2010}.
We sidestep this issue by using a center-independent formulation (\textit{cf.}~\eqref{eqn:density_convolution_spread}); hence, the fact that $\mathbf{c}_{\textsc{dc}}$ is not necessarily continuous with respect to the gauge (as the periodicity implies multiple minimizers) cannot plague $s_{\textsc{dc}}$ even though $\mathbf{s}_{\textsc{dc}} = (\mathcal{S} \rho)(\mathbf{c}_{\textsc{dc}})$ for any $\mathbf{c}_{\textsc{dc}}$ satisfying~\eqref{eqn:density_convolution_center}.

%
%
\begin{figure}[ht!]
\begin{center}
    \includegraphics[width=7.8cm]{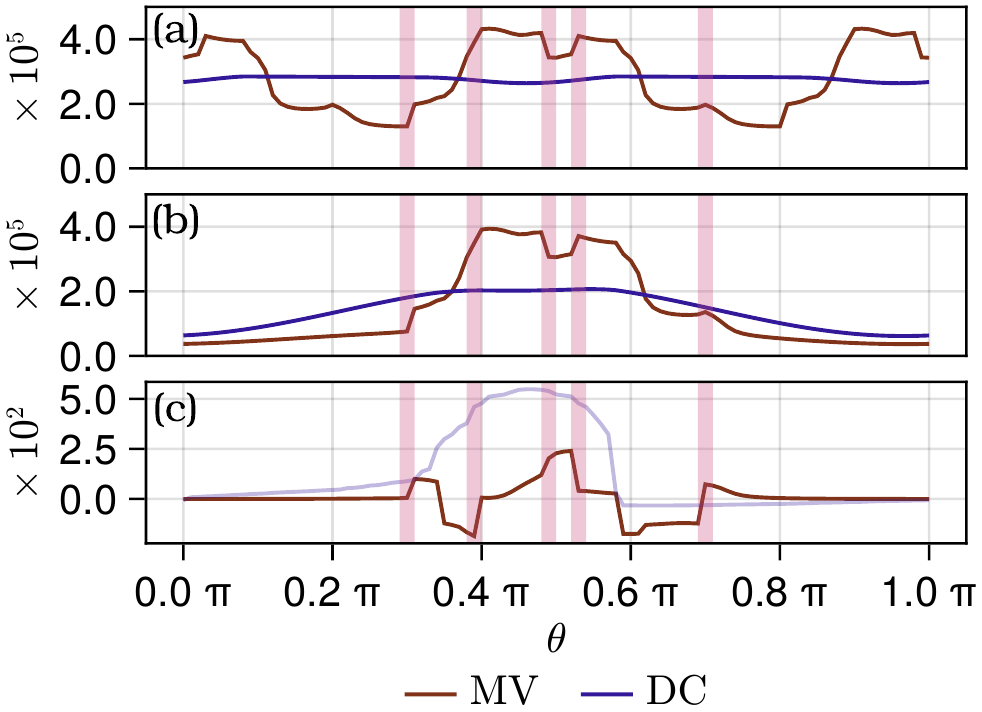}
    \caption{
    For a simple two-band problem with eight $k$-points in 1D, we illustrate that $s_{\textsc{dc}}$ varies continuously with respect to the gauge $U(\theta)$ (which is $\pi$ periodic), while $s_{\textsc{mv}}$ does not. 
    This example contains two Wannier functions $\ket{1}, \ket{2}$ per $k$-point, and we plot (a) the total spread, (b) the spread of $\ket{1}$, and (c) the center of $\ket{1}$ with discontinuities of $s_{\textsc{mv}}$ (and $\mathbf{c}_{\textsc{mv}}$) highlighted~\ref{appendix:two_band}.
}
\label{fig:k_point_discontinuity}
\end{center}
\end{figure}
%
%
To illustrate this advantage, \FIG{Fig.~\ref{fig:k_point_discontinuity}} illustrates how $s_{\textsc{dc}}$ (and $\mathbf{c}_{\textsc{dc}}$) vary with respect to the gauge in a simple two-band problem in 1D.
For comparison, we also consider the Marzari-Vanderbilt (MV) spread ansatz~\cite{MaximallyLocalMarzar1997}, the optimization of which leads to the so-called maximally localized Wannier functions (MLWFs).
\FIG{Fig.~\ref{fig:k_point_discontinuity}(a)} and \FIG{Fig.~\ref{fig:k_point_discontinuity}(b)} show that $s_{\textsc{mv}}$ is discontinuous (both for $\ket{1}$ and the total spread), which can be 
directly traced back to discontinuities in $\mathbf{c}_{\textsc{mv}}$ with respect to the gauge as illustrated in \FIG{Fig.~\ref{fig:k_point_discontinuity}(c)}. 
In contrast, $s_{\textsc{dc}}$ is a smooth function of the gauge even though $\mathbf{c}_{\textsc{dc}}$ is ill-behaved.




The \DC~formulation in~\eqref{eqn:density_convolution_center} and~\eqref{eqn:density_convolution_spread} can be intuitively thought of as \textit{implicitly} using the optimal integration boundary (because a periodic convolution is used) when computing an orbital spread in real space.
Moreover, numerical evaluation of the \DC~integral in~\eqref{eqn:density_convolution} (e.g., via a Fourier Transform) corresponds to a \emph{spectrally} accurate computation of the orbital spread.
This yields a definition that is both consistent in the thermodynamic limit and more accurate for finite-sized domains than common first-order expressions (i.e., as used explicitly by Marzari and Vanderbilt~\cite{MaximallyLocalMarzar1997} via a finite-difference scheme and implicitly by Resta~\cite{QuantumMechaniResta1998} via the use of a single low-frequency Fourier mode when defining the position operator).
Accordingly, $s_{\textsc{dc}}$ is better-suited to quantifying the spread of diffuse orbitals (relative to the unit cell size) and orbitals centered near a unit cell boundary.

%
%
\begin{figure}[t]
    \begin{center}
        \includegraphics[width=.95\linewidth]{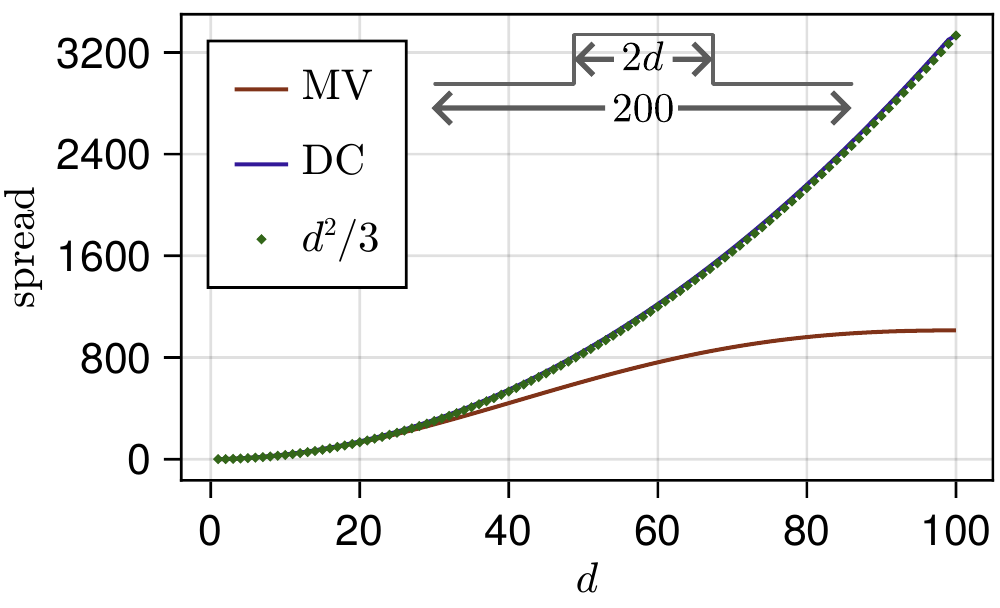}
        \caption{Spread of a square wave in a 1D periodic domain $[-100, 100]$ as a function of the width $2d$. $s_{\textsc{dc}}$ captures the expected $d^2/3$ behavior while $s_{\textsc{mv}}$ only does when $d$ is small relative to the box.
        \label{fig:box}}
    \end{center}
\end{figure}
\begin{figure}[t]
    \begin{center}
        \includegraphics[width=0.9\linewidth]{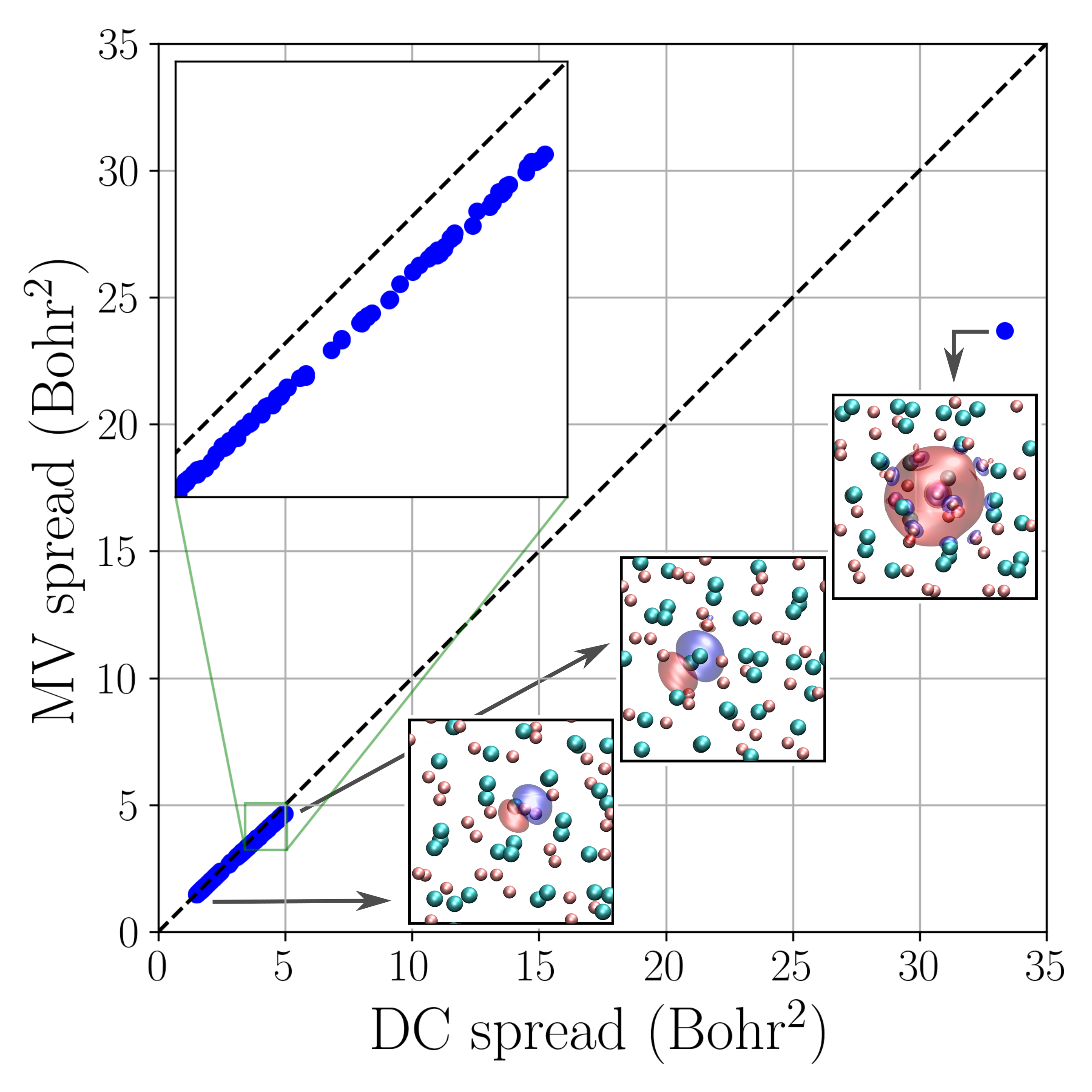}
        \caption{Comparison of $s_{\textsc{dc}}$ and $s_{\textsc{mv}}$ for the localized orbitals in \ce{K33Cl31} computed with \texttt{Wannier90} (blue dots). Insets show selected orbital isosurfaces ($\pm 0.0001$) overlaid with the unit cell (simple cubic; side length $26.59$~Bohr; from Ref.~\cite{OxidationStatePegolo2020}). We observe that $s_{\textsc{mv}}$ deviates from $s_{\textsc{dc}}$ significantly as the orbitals become more diffuse, particularly for the bipolaron state. \label{fig:bipolaron}}
    \end{center}
\end{figure}

%
%
To concretely illustrate the validity of $s_{\textsc{dc}}$ for both local and diffuse orbitals, \FIG{Fig.~\ref{fig:box}} considers the spread of a square wave ($\text{width} = 2d$) in a 1D periodic domain. 
While $s_{\textsc{dc}}$ gives the exact spread ($d^2/3$), $s_{\textsc{mv}}$ is only accurate when $d$ is small relative to the unit cell size and becomes increasingly unreasonable as $d$ grows.  
This discrepancy can also be seen in real systems containing localized orbitals that are relatively diffuse with respect to the unit cell.
A pertinent example is a \ce{K}-doped molten \ce{KCl} salt solution (\ce{K33Cl31})~\cite{LocalizationHSellon1987, ElectronPairinSellon1987, BipolaronsInMFois1988}, in which the bipolaron state is close to the conduction band and quite diffuse. 
\FIG{Fig.~\ref{fig:bipolaron}} compares $s_{\textsc{mv}}$ and $s_{\textsc{dc}}$ and clearly illustrates that $s_{\textsc{mv}}$ deviates from $s_{\textsc{dc}}$ as the orbitals become more diffuse---in this case, severely underestimating the spread of the bipolaron state by roughly 30\%.
%

One of the most practical and prominent uses for an orbital spread expression is within iterative methods for Wannier localization~\cite{MaximallyLocalMarzar2012}. 
While~\eqref{eqn:density_convolution_center} and~\eqref{eqn:density_convolution_spread} can easily be computed given an orbital density, they inherently depend on 
global information (i.e., they are not local operators in $k$-space), which makes $s_{\textsc{dc}}$ cumbersome to optimize directly. 
Hence, we now develop a systematic approximation that is center independent, gauge continuous, and consistent in the thermodynamic limit that can be used as a surrogate for $s_{\textsc{dc}}$ within optimization methods; the final spread of a given orbital (and a center, if desired) can then be computed using~\eqref{eqn:density_convolution_spread} and ~\eqref{eqn:density_convolution_center}.

Our derivation~\ref{appendix:derivation_of_tdc} starts with the truncated
cosine approximation:
\begin{align}
    \mathbf{r}^2 &\gtrapprox \sum_{\mathbf{b}} 2 w_{\mathbf{b}}  \Re \left( 1 - e^{-i \, \mathbf{b}^T \mathbf{r}} \right) ,
    \label{eqn:truncated_cos_approximation}
\end{align}
where $\mathbf{b}$ are selected nearest-neighbor vectors and $w_{\mathbf{b}}$ are the associated weights~\cite{MaximallyLocalMarzar1997}. 
This leads to a lower bound for $s_{\textsc{dc}}$ (that is often tight):
\begin{align}
  \left( \mathcal{S} \rho \right) (\mathbf{r}') &\gtrapprox \sum_{\mathbf{b}} 2 w_{\mathbf{b}} \Re \left( 1 - \hat{\rho}(\mathbf{b}) e^{i \, \mathbf{b}^T \mathbf{r}'} \right) ,
  \label{eqn:decomposition_of_s}
\end{align}
where $\hat{\rho}(\mathbf{b}) = \int_{\mathbb{S}_0} \rho(\mathbf{r}) e^{-i\mathbf{b}^T\mathbf{r}}\mathrm{d}\mathbf{r}$ is the unnormalized Fourier transform of $\rho(\mathbf{r})$.
For Wannier functions, $\hat{\rho}(\mathbf{b}) = \frac{1}{N} \sum_{\mathbf{k}} M_{n,n}^{\mathbf{k}, \mathbf{k} + \mathbf{b}}$, wherein $N$ is the number of electrons and $M$ is the set of $k$-space overlap matrices~\cite{MaximallyLocalMarzar1997}.
We can minimize the right hand side of~\eqref{eqn:decomposition_of_s} by choosing $\mathbf{r}'$ to eliminate the phase of $\hat{\rho}(\mathbf{b})$ yielding:
\begin{align}
  s_{\textsc{dc}} \gtrapprox s_{\textsc{tdc}} \triangleq \sum_{\mathbf{b}} 2 w_{\mathbf{b}} \left( 1 - \abs{\hat{\rho}(\mathbf{b})} \right) ,
  \label{eqn:tdc_spread} 
\end{align}
in which $s_{\textsc{tdc}}$ is the \textit{truncated} \DC~(TDC) approximation (and a formal lower bound) to $s_{\textsc{dc}}$.
In contrast, $s_{\textsc{mv}}$ can either overestimate or underestimate $s_{\textsc{dc}}$ (see \FIG{Fig.~\ref{fig:k_point_discontinuity}}).
%
%
Importantly, $s_{\textsc{tdc}}$ retains the center independence of $s_{\textsc{dc}}$; combined with the fact that $\hat{\rho}(\mathbf{b})$ is continuous with respect to the gauge, this implies that $s_{\textsc{tdc}},$ like $s_{\textsc{dc}},$ is gauge continuous (\FIG{Fig.~\ref{fig:two_band}}).
%
%
%
%

Interestingly,~\eqref{eqn:tdc_spread} has appeared as a generalization of Ref.~\cite{ElectronLocaliResta1999} for finite and $\Gamma$-point systems by Berghold \textit{et al.}~\cite{GeneralAndEffBergho2000}, and in the context of disentanglement by Thygesen \textit{et al.}~\cite{PartlyOccupiedThyges2005} and polarization by Stengel and Spaldin~\cite{AccuratePolariStenge2006}. 
Without the connection to the underlying \DC~formulation presented herein, these seminal works did not fully appreciate the advantages of such an approximation as a \textit{center-independent} and \textit{gauge-continuous} definition of orbital spread, particularly in the iterative localization/optimization context (\textit{vide infra}).
In fact, this formula was often regarded as being equivalent (or essentially equivalent) to $s_{\textsc{mv}}$~\cite{MaximallyLocalMarzar2012}.

To highlight the sizable improvement provided by $s_{\textsc{tdc}}$ for iterative localization, we now compute localized orbitals for a suite of materials using an in-house version of our code~\footnote{Code available at: \url{https://github.com/kangboli/WTP.jl} and  \url{https://github.com/kangboli/SCDM.jl}\label{foot:code}} and the commonly used \texttt{Wannier90} code (version \texttt{3.1.0})~\cite{Wannier90AsAPizzi2020}.
At its core, our code implements the same gradient-based optimization algorithm as \texttt{Wannier90}, but uses $s_{\textsc{tdc}}$ (instead of $s_{\textsc{mv}}$) to define the objective function. 
The TDC scheme introduced herein has a number of favorable properties that allow our code to employ a comparatively simple implementation of manifold optimization in conjunction with standard criteria to reliably determine convergence to localized orbitals.

\begin{figure}[t]
    \begin{center}
        \includegraphics[trim={0.4cm 0.4cm 0.0cm 0.0cm}, clip, width=\linewidth]{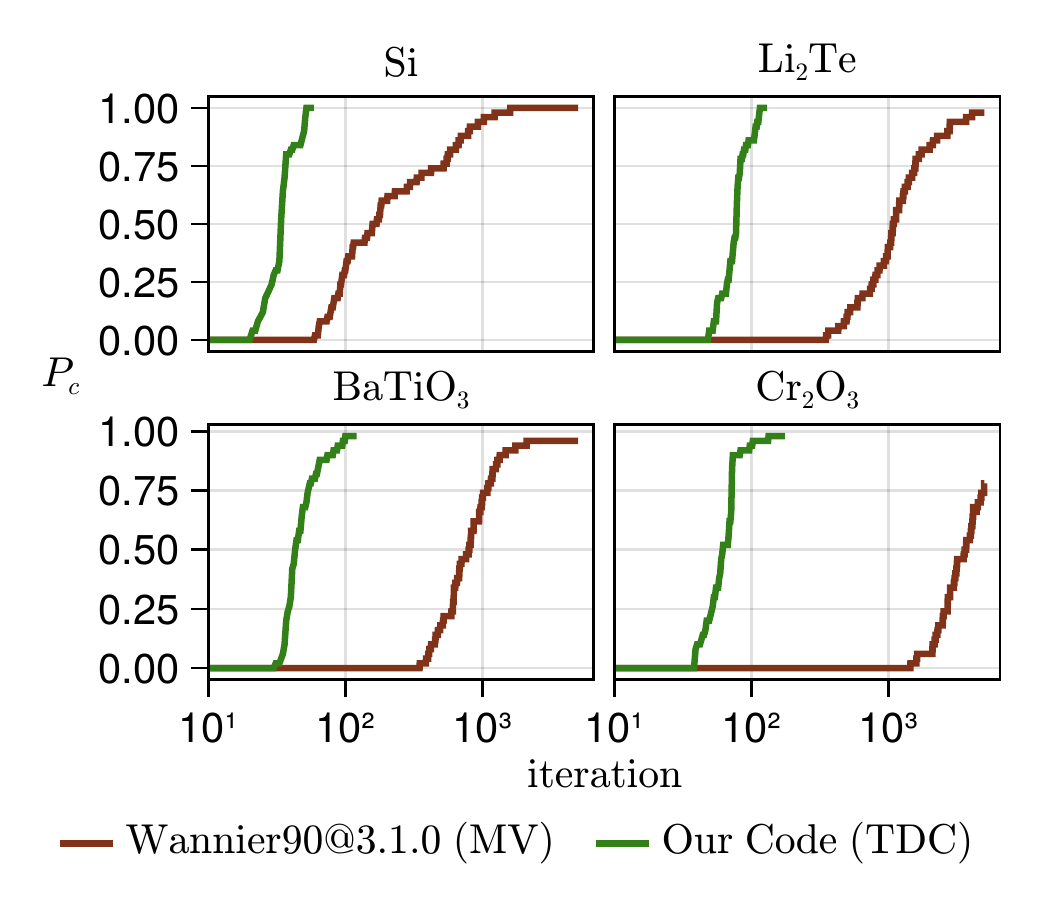}
        \caption{The fraction of orbital localization computations $P_{\rm c}$ that reached success within a given number of iterations for four diverse materials (with success defined retrospectively as reaching within $0.1\%$ of the minimal objective value observed over $50$ runs per material). For fixed $P_{\rm c}$ values, computations using an $s_{\textsc{tdc}}$-based objective function consistently succeeded in $10\times$$-$$70\times$ fewer iterations than those based on $s_{\textsc{mv}}$.
        \label{fig:likelihood}
        }
    \end{center}
\end{figure}

%
To compare the convergence behavior of these two objective functions, we performed a challenging benchmark trial based on four diverse materials: \ce{Si}, \ce{Li2Te}, \ce{BaTiO3}, and \ce{Cr2O3} (\FIG{Table~\ref{tbl:materials}}).
For each material, we performed iterative localization starting from $50$ different randomly generated gauges~\footnote{When using randomly projected $s$-orbitals in \texttt{Wannier90}, the convergence behavior for $s_{\textsc{mv}}$ in \FIG{Fig.~\ref{fig:likelihood}} remains effectively unchanged.}.
\FIG{Fig.~\ref{fig:likelihood}} provides an empirical characterization of the convergence by plotting the fraction of computations $P_{\rm c}$ that succeeded within a given number of iterations (with success defined retrospectively as reaching within $0.1\%$ of the minimal objective value observed over all $50$ runs).
In doing so, we observed that using $s_{\textsc{tdc}}$ consistently reduced the number of iterations required for success by an order of magnitude relative to $s_{\textsc{mv}}$ (i.e., typically $10\times$$-$$70\times$ fewer iterations for a fixed $P_{\rm c}$)~\footnote{Using $s_{\textsc{dc}}$ as an impartial adjudicator, we verified that the ``optimal'' local minima found by these two methods have nearly identical spreads (i.e., to within a couple tenths of a percent)---indicating that success for both codes corresponded to computing orbitals with similar locality.}.
Our $s_{\textsc{tdc}}$-based code is robust and converged in all $200$ cases. 
Moreover, it did not achieve success only twice (once for \ce{BaTiO3} and once for \ce{Cr2O3}), indicating that sub-optimal local minima are---in contrary to common belief---likely rare for these materials.
While \texttt{Wannier90} is often (i.e., by default) run for a fixed number of iterations ($100$ by default), all \texttt{Wannier90} computations performed in this work were allowed to run for a total of $5{,}000$ iterations; $P_{\rm c}$ was then evaluated by retrospectively determining the first iteration when success was achieved for each run.
Even with this favorable criteria, computations with \texttt{Wannier90} often failed---e.g., for \ce{Cr2O3}, \texttt{Wannier90} never reached success within $1{,}000$ iterations and $P_{\rm c} \approx 0.75$ even after $5{,}000$ iterations.
%
%
%
%

%
%
\begin{figure}[t]
    \begin{center}
        \includegraphics[width=\linewidth]{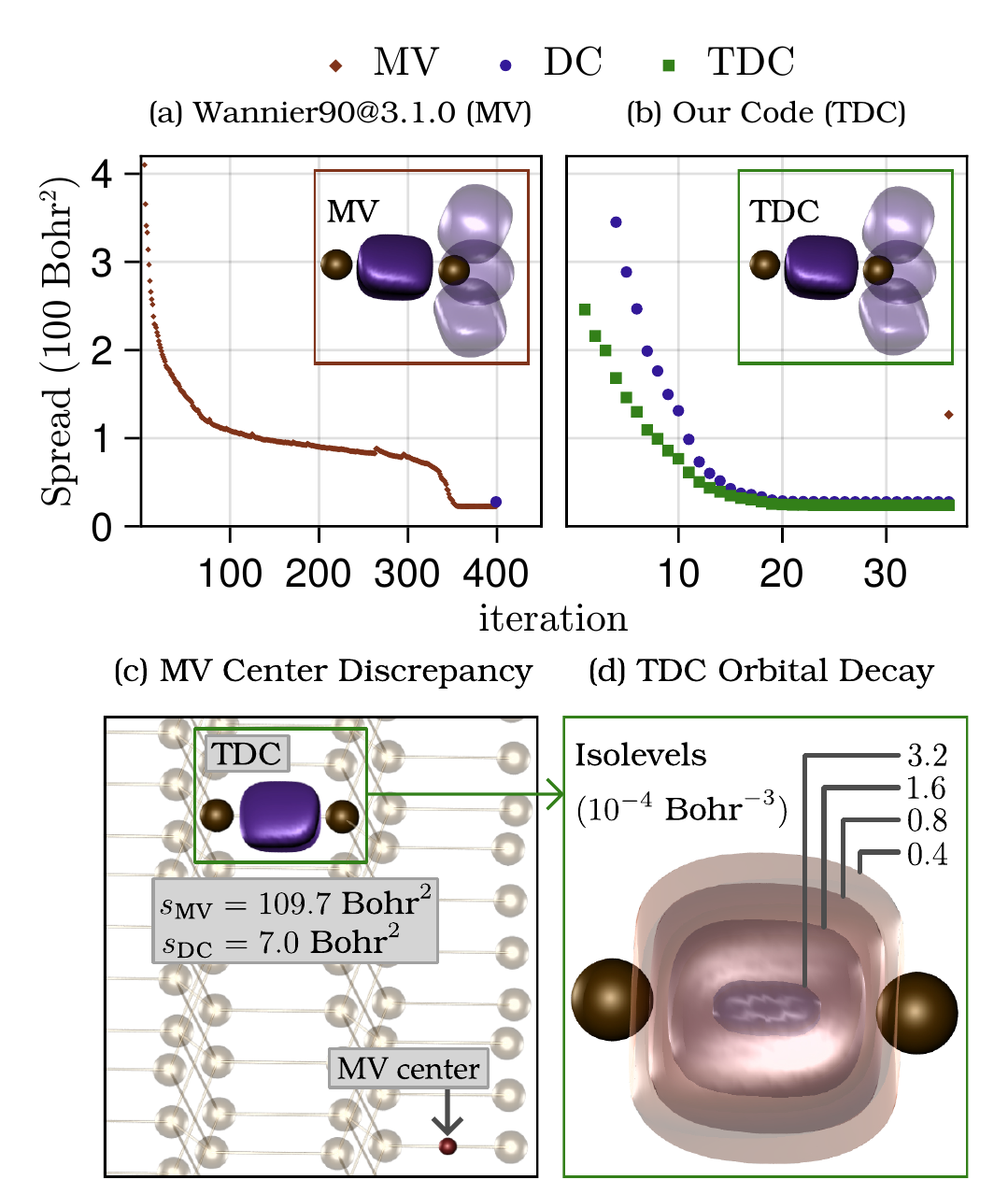}
        \caption{
        Detailed investigation of two successful localization computations for \ce{Si}.
        (a) Non-monotonic decay of $s_{\textsc{mv}}$ for an orbital computed by \texttt{Wannier90} (dark purple orbital in inset).
        (b) Evolution of $s_{\textsc{tdc}}$ for the analogous orbital computed by our code, which achieves success in an order-of-magnitude fewer iterations than \texttt{Wannier90}; upon convergence, $s_{\textsc{mv}}$ severely overestimates the spread of this orbital.
        (c) The distant location of $\mathbf{c}_{\textsc{mv}}$ for the orbital in (b) leads to the large discrepancy between the center-based $s_{\textsc{mv}}$ and $s_{\textsc{tdc}}$ (or $s_{\textsc{dc}}$). 
        (d) Orbital isosurfaces depict the expected (approximate) exponential decay of the orbital in (b) and (c).
        \label{fig:si_trace}
        }
    \end{center}
\end{figure}
%
%
%
To better understand the convergence behavior when using objective functions based on $s_{\textsc{tdc}}$ and $s_{\textsc{mv}}$, \FIG{Fig.~\ref{fig:si_trace}} considers a pair of successful localization computations for \ce{Si}.
\FIG{Fig.~\ref{fig:si_trace}(a)} shows that $s_{\textsc{mv}}$ for one \ce{Si} orbital was not monotonic (e.g., between iterations $250$ and $300$) during the optimization trajectory from \texttt{Wannier90}. 
Such behavior implies that \texttt{Wannier90} encountered (and ``escaped'' from) potentially sub-optimal local minima, consistent with the observations of Canc\`{e}s \textit{et al.}~\cite{RobustDeterminCances2017}; further investigation suggests that these might be non-differentiable (i.e., non-smooth) points where the ``gradient'' (as defined by the standard MV formulation) is non-vanishing---a challenging scenario for any optimizer. 
In contrast, \FIG{Fig.~\ref{fig:si_trace}(b)} shows that $s_{\textsc{tdc}}$ for the analogous orbital monotonically decreases when optimized using our code, which reaches the same localized orbitals (per a visual metric, see inset) as \texttt{Wannier90} in an order-of-magnitude fewer iterations. 
Matching our theoretical expectations, $s_{\textsc{tdc}}$ is a lower bound for $s_{\textsc{dc}}$ (which is easily computed for any fixed gauge), with these metrics converging as the orbital becomes more localized.
However, $s_{\textsc{mv}}$ severely overestimates the spread of the highlighted localized orbital computed using our code.
This discrepancy is due to the distant location of the (necessarily) computed orbital center as seen in \FIG{Fig.~\ref{fig:si_trace}(c)}---highlighting a potential danger of a center-based definition  such as $s_{\textsc{mv}}$. 
Consistent with both $s_{\textsc{tdc}}$ and $s_{\textsc{dc}}$, \FIG{Fig.~\ref{fig:si_trace}(d)} confirms that our code did indeed compute a local orbital exhibiting exponential decay. 
%
%
As illustrated by \FIG{Fig.~\ref{fig:si_trace}(b)-(d)} the $s_{\textsc{mv}}$ metric can sometimes fail to recognize localized orbitals, $s_{\textsc{tdc}}$ is not prone to such issues by construction.
%

%
We have presented the \DC~formulation for the spread (and center) of an orbital under PBCs and illustrated its clear theoretical advantages (e.g., gauge continuity) relative to prior work. 
Moreover, preliminary evidence strongly suggests that $s_{\textsc{tdc}}$ (a systematic approximation to $s_{\textsc{dc}}$) is favorable for use in iterative localization methods---exhibiting significantly improved performance and robustness in challenging scenarios relative to prior work. 
While we have only discussed insulating systems in this work, entangled systems can also be treated by adapting Ref.~\cite{VariationalForDamle2019} to use $s_{\textsc{tdc}}$ and is therefore a potentially interesting direction for future research. 
Accompanying this work are two Julia packages, \texttt{WTP.jl} and \texttt{SCDM.jl}~\cite{Note1}, that allow researchers to experiment with and expand upon the ideas presented herein.

%

\vspace{1em}

\begin{acknowledgements}
\noindent \textit{All authors acknowledge Antoine Levitt for helpful scientific discussions and feedback on an early version of this manuscript. This material is based upon work supported by the National Science Foundation under Grant No.\ CHE-1945676. RAD also gratefully acknowledges financial support from an Alfred P.\ Sloan Research Fellowship.}
\end{acknowledgements}

\bibliography{lib,more_refs} 
\clearpage

\onecolumngrid

\appendix

\renewcommand{\thesection}{\Roman{section}}
\section{Derivation of the \TDC~spread}
\label{appendix:derivation_of_tdc}

We provide a more detailed derivation of the \TDC~spread and prove that it
is a lower bound to the \DC~spread. First, we make a truncated cosine
approximation
\begin{align}
    \mathbf{r}^2 = \sum_{\mathbf{b}} w_{\mathbf{b}} (\mathbf{b}^T \mathbf{r})^2 
    &\gtrapprox \sum_{\mathbf{b}} 2 w_{\mathbf{b}}  \Re (1 - e^{-i \mathbf{b}^T \mathbf{r}}), \label{eqn:truncated_cos_approximation_supplement}
\end{align}
where $\mathbf{b}$ are discrete k-points~\footnote{
The Brillouin zone lattice points are  integer combinations 
of Brillouin zone basis vectors $\mathbf{b} = \sum_i^n n_i \mathbf{b}_i$. The basis vectors are integer (number of k-points along each direction) fractions 
of the reciprocal lattice basis vectors.
}  chosen in the neighborhood
of the Gamma point (details in~\cite{Wannier90AToMostof2008}),
and they satisfy the identity:
$\sum_{\mathbf{b}} w_{\mathbf{b}} \mathbf{b} \mathbf{b}^T = I$~\cite{MaximallyLocalMarzar1997}.
Notably, the approximation~\eqref{eqn:truncated_cos_approximation_supplement} 
underestimates $\mathbf{r}^2$, so applying it to $\mathcal{S} \rho$ 
(the \DC) gives us an underestimate, which we denote as 
$\mathcal{T}\rho$ (the \TDC)
\begin{align}\label{eqn:decomposition_of_s_supplement}
        (\mathcal{S} \rho) (\mathbf{r}') 
          &\gtrapprox \sum_{\mathbf{b}} 2 w_{\mathbf{b}} \Re (1 - \hat{\rho}(\mathbf{b}) e^{i \mathbf{b}^T \mathbf{r}'}) \triangleq (\mathcal{T}\rho)(\mathrm{r}'),\\
          \text{where }
        \hat{\rho} (\mathbf{b}) &= \int_{\mathbb{S}_{\mathbf{r}'}}
        \rho(\mathbf{r}) e^{-i \mathbf{b}^T \mathbf{r}} \mathrm{d} \mathbf{r}
        = \int_{\mathbb{S}_{\mathbf{0}}}
         \rho(\mathbf{r}) e^{-i \mathbf{b}^T \mathbf{r}} \mathrm{d} \mathbf{r}.
         \label{eqn:FT_of_rho}
\end{align}
Conveniently, $\hat{\rho}$ is just 
the unnormalized Fourier transform of $\rho$, in which
the shift of the integration boundary is justified
because the integrand is periodic when $\mathbf{b}$ are discrete k-points.

The \TDC~in Eq.~\eqref{eqn:decomposition_of_s_supplement} does not have an 
apparent analytical minimum, but each term in the sum does have one, 
and they give us a lower bound on the minimum of the \TDC.
\begin{align}
    \min_{\mathbf{r}'} (\mathcal{T}\rho) (\mathbf{r}') 
    &\gtrapprox \sum_{\mathbf{b}} \min_{\mathbf{r}'} \left(2w_{\mathbf{b}} \Re (1-\hat{\rho}(\mathbf{b}) e^{i \mathbf{b}^T \mathbf{r}'})\right) 
    =  \sum_{\mathbf{b}} 2 w_{\mathbf{b}} (1 - \abs{\hat{\rho}(\mathbf{b})}).
    \label{eqn:tdc_spread_supplement}
\end{align}
The lower bound is attained if each term in the sum is 
independently minimized, which corresponds to 
$\mathbf{b}^T \mathbf{r}' = -\Im \ln \hat{\rho}(\mathbf{b})$.
This is a linear system that has a solution for simple lattices, but  
it can be overdetermined for more complex lattices, so 
the bound is not always attainable. In either case, we can find a least
square solution with the identity:
$\sum_{\mathbf{b}} w_{\mathbf{b}} \mathbf{b} \mathbf{b}^T = I$, 
which gives us the \TDC~center
\begin{equation}\label{eqn:tdc_center}
    \mathbf{c}_{\textsc{tdc}} \triangleq -\sum_{\mathbf{b}} w_{\mathbf{b}}
    \mathbf{b} \Im \ln \hat{\rho}(\mathbf{b}). 
\end{equation}
Formally, both the \TDC~center and the \TDC~spread 
are accurate only for well-localized orbitals, 
in which case the logarithm is stable and 
the least square procedure has a small residual. 

\section{Density of the Wannier Functions}
\label{appendix:density_of_wannier}

In DFT, the eigenstates of a periodic Hamiltonian 
take the form of the Bloch orbitals
\begin{equation}
    \psi_{n, \mathbf{k}}(\mathbf{r}) = u_{n, \mathbf{k}}(\mathbf{r}) e^{i \mathbf{k}^T \mathbf{r}},
\end{equation}
where $\mathbf{k}$ is a $k$-point in the first Brillouin zone 
and $u_{n, \mathbf{k}}(\mathbf{r}) =
u_{n, \mathbf{k}}(\mathbf{r} + \mathbf{R})$ is a
function periodic under lattice translations by $\mathbf{R}$.
In terms of the Bloch orbitals, the density of a 
Wannier function $\rho_n$ can be written as 
\begin{align}
    \rho_n(\mathbf{r}) &= \frac{1}{N}
    \sum_{\mathbf{k}, \mathbf{k}'} e^{i (\mathbf{k}'-\mathbf{k})^T \mathbf{r}} v_{n, \mathbf{k}}^* (\mathbf{r}) v_{n, \mathbf{k}'} (\mathbf{r}) \mathrm{d} \mathbf{r},\label{eqn:wannier_density}
\end{align}
where $N$ is the number of $k$-points sampled in the 
first Brillouin zone, and the periodic part 
$u_{n, \mathbf{k}}$ has been transformed by a gauge  
$v_{n, \mathbf{k}}(\mathbf{r}) = \sum_m U_{m, n}^{\mathbf{k}} u_{m, \mathbf{k}}(\mathbf{r})$. 
 The unnormalized Fourier transform of $\rho_n$ is then
\begin{equation}\label{eqn:neighbor_integrals}
    \begin{aligned}
    \hat{\rho}_n(\mathbf{b}) &= \frac{1}{N} \sum_{\mathbf{k}, \mathbf{k}'}
\int_{\mathbb{S}_{\mathbf{0}}} 
e^{i (\mathbf{k}'-\mathbf{k}-\mathbf{b})^T \mathbf{r}} v_{n, \mathbf{k}}^*(\mathbf{r}) v_{n, \mathbf{k}'}(\mathbf{r}) \mathrm{d} \mathbf{r} 
= \frac{1}{N} \sum_{\mathbf{k}} \int_{\mathbb{S}_{\mathbf{0}}} v_{n, \mathbf{k}}^* (\mathbf{r}) v_{n, \mathbf{k}+\mathbf{b}}(\mathbf{r}) \mathrm{d} \mathbf{r}.
    \end{aligned}
\end{equation}
It is possible for the vector $\mathbf{k} + \mathbf{b}$ in~\eqref{eqn:neighbor_integrals} to be out of the first Brillouin zone. In that case, our notation is to be interpreted as
\begin{equation}
    v_{n, \mathbf{k} + \mathbf{b}}(\mathbf{r}) \triangleq \exp(i \mathbf{G}^T \mathbf{r}) v_{n, \mathbf{k} + \mathbf{b} + \mathbf{G}}(\mathbf{r}), \label{eqn:outside_interpretation}
\end{equation}
where $\mathbf{G}$ is the reciprocal lattice vector s.t. $\mathbf{k} + \mathbf{b} + \mathbf{G}$ folds into the first Brillouin zone.

Now, using MLWF notations, the unnormalized Fourier transform becomes
\begin{align}
\hat{\rho}_n(\mathbf{b}) &= \frac{1}{N} \sum_{\mathbf{k}} M_{n,n}^{\mathbf{k}, \mathbf{k} + \mathbf{b}}\label{eqn:unification},\\
    \text{where } 
    M_{m,n}^{\mathbf{k}, \mathbf{k} + \mathbf{b}} = 
    \bra{v_{m, \mathbf{k}}} \ket{v_{n,\mathbf{k} + \mathbf{b}}} 
    &=\sum_{p,q} U_{p,m}^{\mathbf{k} *}
    S_{p,q}^{\mathbf{k}, \mathbf{k} + \mathbf{b}} 
    U^{\mathbf{k} + \mathbf{b}}_{q,n},
    S_{m,n}^{\mathbf{k}, \mathbf{k} + \mathbf{b}} = 
    \bra{u_{m, \mathbf{k}}} \ket{u_{n,\mathbf{k} + \mathbf{b}}}.
\end{align}
This formula is only notationally different from what Thygesen et al.~\cite{PartlyOccupiedThyges2005} derived.

\section{The Gradient and Manifold Optimization}
\label{appendix:gradient}

Conventionally, to minimize the total spread with respect to the gauge, 
one derives a \textit{constrained gradient} of the total spread by 
looking at the effect of adding a constrained first order 
perturbation to $U^{\mathbf{k}}$. One then applies a gradient descent
algorithm that is slightly tweaked to work with the 
constrained gradient~\cite{MaximallyLocalMarzar1997, PartlyOccupiedThyges2005}.
Although this approach is correct, it does not easily fit into 
a generic differentiation and optimization framework, making it 
difficult both to understand and to leverage existing tools.

By constrast, we derive the \textit{Euclidean gradient}  
(i.e. unconstrained) of the total \TDC~spread
\begin{equation}\label{eqn:total_spread}
    \Omega_{\textsc{tdc}}(U) = \sum_{n, \mathbf{b}} 2 w_{\mathbf{b}} (1- |\hat{\rho}_n(\mathbf{b})|),
\end{equation}
and apply the (Stiefel) manifold optimization.
Our description will be procedurally equivalent 
to the conventional one,
but our approach can reach a broader audience 
and enable the use of automatic differentiation and 
manifold optimization tools.

To derive the (unconstrained) gradient with respect to 
$U^{\mathbf{k}}$, we first split $U^{\mathbf{k}}$ and 
$S^{\mathbf{k}, \mathbf{k}+\mathbf{b}}$ into their 
real and imaginary parts: $U^{\mathbf{k}} = 
X^{\mathbf{k}} + i Y^{\mathbf{k}}$ and
$S^{\mathbf{k}, \mathbf{k}+\mathbf{b}} = P^{\mathbf{k}, \mathbf{k}+\mathbf{b}} + i Q^{\mathbf{k},
\mathbf{k}+\mathbf{b}}$.
$M_{n,n}^{\mathbf{k},\mathbf{k}+\mathbf{b}}$ can then
be expressed as
\begin{align}
    M_{n,n}^{\mathbf{k},\mathbf{k}+\mathbf{b}} =&
     \sum_{r,s} \left(X_{r,n}^{\mathbf{k}} - iY_{r,n}^{\mathbf{k}}\right)
    \left(P_{r,s}^{\mathbf{k}, \mathbf{k} + \mathbf{b}} + iQ_{r,s}^{\mathbf{k}, \mathbf{k} + \mathbf{b}}\right)
    \left(X_{s,n}^{\mathbf{k} + \mathbf{b}} + iY_{s,n}^{\mathbf{k} + \mathbf{b}}\right)
    = C_{n,n}^{\mathbf{k},\mathbf{k}+\mathbf{b}} + i D_{n,n}^{\mathbf{k},\mathbf{k}+\mathbf{b}},\\
    C_{n,n}^{\mathbf{k},\mathbf{k}+\mathbf{b}} =&
    \sum_{r,s} \left(
    P_{r,s}^{\mathbf{k}, \mathbf{k} + \mathbf{b}}
    (X_{r,n}^{\mathbf{k}} X_{s,n}^{\mathbf{k} + \mathbf{b}} + 
    Y_{r,n}^{\mathbf{k}} Y_{s,n}^{\mathbf{k} + \mathbf{b}})
    - Q_{r,s}^{\mathbf{k}, \mathbf{k} + \mathbf{b}}(
X_{r,n}^{\mathbf{k}}Y_{s,n}^{\mathbf{k} + \mathbf{b}}
- Y_{r,n}^{\mathbf{k}} X_{s,n}^{\mathbf{k} + \mathbf{b}}) \right),\label{eqn:c_real_imag}\\
    D_{n,n}^{\mathbf{k},\mathbf{k}+\mathbf{b}} = &
    \sum_{r,s}\left(
    Q_{r,s}^{\mathbf{k}, \mathbf{k} + \mathbf{b}}
    (X_{r,n}^{\mathbf{k}} X_{s,n}^{\mathbf{k} + \mathbf{b}} +
    Y_{r,n}^{\mathbf{k}} Y_{s,n}^{\mathbf{k} + \mathbf{b}}) 
    + P_{r,s}^{\mathbf{k}, \mathbf{k} + \mathbf{b}}(
X_{r,n}^{\mathbf{k}}Y_{s,n}^{\mathbf{k} + \mathbf{b}}
- Y_{r,n}^{\mathbf{k}} X_{s,n}^{\mathbf{k} + \mathbf{b}}) \right).\label{eqn:d_real_imag}
\end{align}
The nontrivial part of our objective function and its derivative can be rewritten as
\begin{align}
    \sum_n|\hat{\rho}_n(\mathbf{b})| &=
    \sum_n \frac{1}{N}\sqrt{ 
        \left(\sum_{\mathbf{k}} C_{n,n}^{\mathbf{k}, \mathbf{k} + \mathbf{b}}\right)^2 +
        \left(\sum_{\mathbf{k}} D_{n,n}^{\mathbf{k}, \mathbf{k} + \mathbf{b}}\right)^2
},\\
        \sum_n \frac{\partial |\hat{\rho}_n(\mathbf{b})| }{\partial U_{i,j}^{\mathbf{k}}} 
        &= \sum_{n, \mathbf{h}} 
        \frac{\partial |\hat{\rho}_n(\mathbf{b})|}{\partial C_{n,n}^{\mathbf{h}, \mathbf{h} + \mathbf{b}}} 
    \left( 
\frac{\partial C_{n,n}^{\mathbf{h}, \mathbf{h} + \mathbf{b}}}{\partial X_{i,j}^{\mathbf{k}}} + 
i\frac{\partial C_{n,n}^{\mathbf{h}, \mathbf{h} + \mathbf{b}}}{\partial Y_{i,j}^{\mathbf{k}}}
\right) +
 \frac{\partial |\hat{\rho}_n(\mathbf{b})|}{\partial D_{n,n}^{\mathbf{h}, \mathbf{h} + \mathbf{b}}} 
\left(
    \frac{\partial D_{n,n}^{\mathbf{h}, \mathbf{h} + \mathbf{b}}}
    {\partial X_{i,j}^{\mathbf{k}}} + 
    i \frac{\partial D_{n,n}^{\mathbf{h}, \mathbf{h} + \mathbf{b}}}
    {\partial Y_{i,j}^{\mathbf{k}}}
\right).
\end{align}
The derivative of $C$ with respect to $X$ and $Y$ are
\begin{align}
    \frac{\partial C_{n,n}^{\mathbf{h}, \mathbf{h} + \mathbf{b}}}{\partial X_{i,j}^{\mathbf{k}}}  
    =& \sum_{r,s} P_{r,s}^{\mathbf{h}, \mathbf{h} + \mathbf{b}}
    (\delta_{i,j,\mathbf{k}}^{r,n,\mathbf{h}} X_{s,n}^{\mathbf{h} + \mathbf{b}} +
    X_{r,n}^{\mathbf{h}} \delta_{i,j,\mathbf{k}}^{s,n, \mathbf{h} + \mathbf{b}})
    - Q_{r,s}^{\mathbf{h}, \mathbf{h} + \mathbf{b}}(
    \delta_{i,j,\mathbf{k}}^{r,n, \mathbf{h}}Y_{s,n}^{\mathbf{h} + \mathbf{b}}
    - Y_{r,n}^{\mathbf{h}} \delta_{i,j,\mathbf{k}}^{s,n, \mathbf{h} + \mathbf{b}}),\\
    \frac{\partial C_{n,n}^{\mathbf{h}, \mathbf{h} + \mathbf{b}}}{\partial Y_{i,j}^{\mathbf{k}}}  
    =& \sum_{r,s} P_{r,s}^{\mathbf{h}, \mathbf{h} + \mathbf{b}}
    (\delta_{i,j,\mathbf{k}}^{r,n,\mathbf{h}} Y_{s,n}^{\mathbf{h} + \mathbf{b}} +
    Y_{r,n}^{\mathbf{h}} \delta_{i,j,\mathbf{k}}^{s,n, \mathbf{h} + \mathbf{b}})
    - Q_{r,s}^{\mathbf{h}, \mathbf{h} + \mathbf{b}}(
    X_{r,n}^{\mathbf{h}} \delta_{i,j,\mathbf{k}}^{s,n, \mathbf{h} + \mathbf{b}}
    -\delta_{i,j,\mathbf{k}}^{r,n, \mathbf{h}} X_{s,n}^{\mathbf{h} + \mathbf{b}}).
\end{align}
Combining them gives
\begin{equation}
    \begin{aligned}
\frac{\partial C_{n,n}^{\mathbf{h}, \mathbf{h} + \mathbf{b}}}{\partial U_{i,j}^{\mathbf{k}}} 
= & 
    \sum_{r,s} \left(P_{r,s}^{\mathbf{h}, \mathbf{h} + \mathbf{b}}
    (\delta_{i,j,\mathbf{k}}^{r,n,\mathbf{h}} U_{s,n}^{\mathbf{h} + \mathbf{b}} +
    U_{r,n}^{\mathbf{h}} \delta_{i,j,\mathbf{k}}^{s,n, \mathbf{h} + \mathbf{b}}) 
    - i Q_{r,s}^{\mathbf{h}, \mathbf{h} + \mathbf{b}}(
    U_{r,n}^{\mathbf{h}} \delta_{i,j,\mathbf{k}}^{s,n, \mathbf{h} + \mathbf{b}}
    -\delta_{i,j,\mathbf{k}}^{r,n, \mathbf{h}} U_{s,n}^{\mathbf{h} + \mathbf{b}})\right).
    \end{aligned}
\end{equation}

Before we continue, we introduce some tensor symmetries 
that will simplify our result. 
Because $S^{\mathbf{k}, \mathbf{k}+\mathbf{b}}_{m,n} =
\bra{u_m^{\mathbf{k}}} \ket{u_n^{\mathbf{k}+\mathbf{b}}} =
S^{\mathbf{k}+\mathbf{b}, \mathbf{k} *}_{n,m}$, we have the symmetry
\begin{align}
    P^{\mathbf{k}, \mathbf{k}+\mathbf{b}}_{m,n} = P^{\mathbf{k}+\mathbf{b}, \mathbf{k}}_{n, m}, \quad Q^{\mathbf{k}, \mathbf{k}+\mathbf{b}}_{m,n} = -Q^{\mathbf{k}+\mathbf{b}, \mathbf{k}}_{n, m}. \label{eqn:symmetry_pq}
\end{align}
The same symmetry applies to $C$ and $D$ for similar reasons.
Another symmetry comes from $\rho_n(\mathbf{r})$ being real,
which leads to 
$\hat{\rho}_n(-\mathbf{b}) = \hat{\rho}^{*}_n(\mathbf{b})$.
These symmetries allow as to write the derivative through $C$ as 
\begin{equation}
    \begin{aligned}
        \sum_{n, \mathbf{h}, \mathbf{b}} 
        \omega_{\mathbf{b}} \frac{\partial |\hat{\rho}_n(\mathbf{b})|}{\partial C_{n,n}^{\mathbf{h}, \mathbf{h} + \mathbf{b}}} 
        \frac{\partial C_{n,n}^{\mathbf{h}, \mathbf{h} + \mathbf{b}}}{\partial U_{i,j}^{\mathbf{k}}}  
=
        &\sum_{\mathbf{b}} \omega_{\mathbf{b}}
        \left(
        \frac{\partial |\hat{\rho}_j(\mathbf{b})|}{\partial C_{j,j}^{\mathbf{k}, \mathbf{k} + \mathbf{b}}} 
        \sum_{s} P_{i,s}^{\mathbf{k}, \mathbf{k} + \mathbf{b}}
        U_{s,j}^{\mathbf{k} + \mathbf{b}} +
        \frac{\partial |\hat{\rho}_j(\mathbf{b})|}{\partial C_{j,j}^{\mathbf{k} - \mathbf{b}, \mathbf{k}}}
        \sum_{r} P_{r,i}^{\mathbf{k} - \mathbf{b}, \mathbf{k}}
        U_{r,j}^{\mathbf{k}-\mathbf{b}} \right.\\
        &\left.-
        i\frac{\partial |\hat{\rho}_j(\mathbf{b})|}{\partial C_{j,j}^{\mathbf{k} - \mathbf{b}, \mathbf{k}}}
        \sum_{r} Q_{r,i}^{\mathbf{k} - \mathbf{b}, \mathbf{k} }
        U_{r,j}^{\mathbf{k} - \mathbf{b}} +
        i \frac{\partial |\hat{\rho}_j(\mathbf{b})|}{\partial C_{j,j}^{\mathbf{k}, \mathbf{k} + \mathbf{b}}} 
        \sum_s Q_{i,s}^{\mathbf{k}, \mathbf{k} + \mathbf{b}}
        U_{s,j}^{\mathbf{k} + \mathbf{b}} \right)\\ 
        =&2\sum_{\mathbf{b}} \omega_{\mathbf{b}}
        \frac{\partial |\hat{\rho}_j(\mathbf{b})|}{\partial C_{j,j}^{\mathbf{k}, \mathbf{k} + \mathbf{b}}} 
        \sum_s \left(
            P_{i,s}^{\mathbf{k}, \mathbf{k} + \mathbf{b}}
            U_{s,j}^{\mathbf{k} + \mathbf{b}} +
            i Q_{i,s}^{\mathbf{k}, \mathbf{k} + \mathbf{b}}
            U_{s,j}^{\mathbf{k} + \mathbf{b}} \right)\\
        =&2\sum_{\mathbf{b}} \omega_{\mathbf{b}}
        \frac{\partial |\hat{\rho}_j(\mathbf{b})|}{\partial C_{j,j}^{\mathbf{k}, \mathbf{k} + \mathbf{b}}} 
        \sum_s 
            S_{i,s}^{\mathbf{k}, \mathbf{k} + \mathbf{b}}
            U_{s,j}^{\mathbf{k} + \mathbf{b}},
    \end{aligned}
\end{equation}
where we have replaced $-\mathbf{b}$ with $\mathbf{b}$ because the sum is symmetric $(\sum_{\mathbf{b}} = \sum_{-\mathbf{b}})$.
Similarly, the derivative through $D$ can be derived  as
\begin{equation}
    \begin{aligned}
        \sum_{n, \mathbf{h}, \mathbf{b}} \omega_{\mathbf{b}}
        \frac{\partial |\hat{\rho}_n(\mathbf{b})|}{\partial D_{n,n}^{\mathbf{h}, \mathbf{h} + \mathbf{b}}} 
        \frac{\partial D_{n,n}^{\mathbf{h}, \mathbf{h} + \mathbf{b}}}{\partial U_{i,j}^{\mathbf{k}}}  
=
        &\sum_{\mathbf{b}} \omega_{\mathbf{b}}
        \left(\frac{\partial |\hat{\rho}_j(\mathbf{b})|}{\partial D_{j,j}^{\mathbf{k}, \mathbf{k} + \mathbf{b}}} 
        \sum_{s} Q_{i,s}^{\mathbf{k}, \mathbf{k} + \mathbf{b}}
        U_{s,j}^{\mathbf{k} + \mathbf{b}} +
        \frac{\partial |\hat{\rho}_j(\mathbf{b})|}{\partial D_{j,j}^{\mathbf{k} - \mathbf{b}, \mathbf{k}}}
        \sum_{r} Q_{r,i}^{\mathbf{k} - \mathbf{b}, \mathbf{k}}
        U_{r,j}^{\mathbf{k}-\mathbf{b}} \right.\\
        &\left.+
        i\frac{\partial |\hat{\rho}_j(\mathbf{b})|}{\partial D_{j,j}^{\mathbf{k} - \mathbf{b}, \mathbf{k}}}
        \sum_{r} P_{r,i}^{\mathbf{k} - \mathbf{b}, \mathbf{k} }
        U_{r,j}^{\mathbf{k} - \mathbf{b}} -
        i \frac{\partial |\hat{\rho}_j(\mathbf{b})|}{\partial D_{j,j}^{\mathbf{k}, \mathbf{k} + \mathbf{b}}} 
        \sum_s P_{i,s}^{\mathbf{k}, \mathbf{k} + \mathbf{b}}
        U_{s,j}^{\mathbf{k} + \mathbf{b}} \right) \\ 
        =&2\sum_{\mathbf{b}}\omega_{\mathbf{b}}
        \frac{\partial |\hat{\rho}_j(\mathbf{b})|}{\partial D_{j,j}^{\mathbf{k}, \mathbf{k} + \mathbf{b}}} 
        \sum_s \left(
            -iP_{i,s}^{\mathbf{k}, \mathbf{k} + \mathbf{b}}
            U_{s,j}^{\mathbf{k} + \mathbf{b}} +
            Q_{i,s}^{\mathbf{k}, \mathbf{k} + \mathbf{b}}
            U_{s,j}^{\mathbf{k} + \mathbf{b}} \right)\\
        =&-2i\sum_{\mathbf{b}} \omega_{\mathbf{b}}
        \frac{\partial |\hat{\rho}_j(\mathbf{b})|}{\partial D_{j,j}^{\mathbf{k}, \mathbf{k} + \mathbf{b}}} 
        \sum_s 
            S_{i,s}^{\mathbf{k}, \mathbf{k} + \mathbf{b}}
            U_{s,j}^{\mathbf{k} + \mathbf{b}}.
    \end{aligned}
\end{equation}
Finally, the full expression becomes 
\begin{equation}
    G_{i,j}^{\mathbf{k}} = -2\sum_{n,\mathbf{b}} \omega_{\mathbf{b}} \frac{\partial \abs{\hat{\rho}_n(\mathbf{b})}}{\partial U_{i,j}^{\mathbf{k}}} = - \frac{4}{N}  \sum_{\mathbf{b}, s} \omega_{\mathbf{b}} \frac{\hat{\rho}_j^*(\mathbf{b})}{\abs{\hat{\rho}_j(\mathbf{b})}}  S_{i,s}^{\mathbf{k}, \mathbf{k}+\mathbf{b}} U_{s, j}^{\mathbf{k}+\mathbf{b}}.
\end{equation}


Given the gradient $G^{\mathbf{k}}$, any constrained 
optimization scheme should  work. 
We will use the manifold optimization ~\cite{OptimizationAlAbsil2007,
ABriefIntroduHuJi2020, Bergmann2022}, which requires,
in addition to the unconstrained gradient, 
the tangent plane projection and a manifold retraction. 
In particular, the gauge is constrained on the (power) Stiefel manifold $\mathcal{M}^N$~\cite{
ConjugateGradiAbruda2009,  OptimizationAlManton2002, TheGeometryOfEdelma1998}, where
 $\mathcal{M} = \left\{A \in \mathbb{C}^{N_b \times N_b} : A^{\dagger} A = I\right\}$, and $N_b$ is the number of bands. The projection onto $\mathcal{M}^N$ is an
antisymmetrization $(G^{\mathbf{k}} - G^{\mathbf{k} \dagger})/2$, and the retraction we use is
 \begin{equation}
    \mathrm{rt}(U^{\mathbf{k}}, \Delta U^{\mathbf{k}})
    = U^{\mathbf{k}} \exp(U^{\mathbf{k}\dagger} \Delta U^{\mathbf{k}}),
\end{equation}
where $\Delta U$ is the projected gradient step $\alpha \left(G^{\mathbf{k}} - G^{\mathbf{k}\dagger}\right)/2$
with step size $\alpha$. 
The choice of the retraction is not unique as long as
it agrees with $U + \Delta U$ to linear order. 
We chose our retraction  to procedurally reproduce 
the same optimization algorithm as \texttt{Wannier90}, and
it does agree with $U + \Delta U$ to linear order 
\begin{equation}
    U^\mathbf{k}\exp(U^{\mathbf{k} \dagger} \Delta U^{\mathbf{k}})
    \approx U^{\mathbf{k}} (1 + U^{\mathbf{k}\dagger} \Delta U^{\mathbf{k}})
    = U^{\mathbf{k}} + \Delta U^{\mathbf{k}}.
\end{equation}


\section{Materials}
\label{appendix:materials}


We provide a list of materials with the number of k-points and bands 
in Table.~\ref{tbl:materials}

\begin{table}[ht]
\begin{center}
\begin{tabular}{| l | r | r |}
\hline
    Material & \# k-points & \# bands\\\hline
    \ce{Si} & $64$ & $4$\\
    \ce{BaTiO3} & $64$ & $9$ \\
    \ce{Li2Te} & $125$ & $6$ \\
    \ce{Cr2O3} & $216$ & $6$\\\hline
\end{tabular}
\caption{\label{tbl:materials}
    List of materials: \ce{Li2Te} has been previously reported to fail for
    Wannier90 with random start. \ce{Cr2O3} is chosen because it is suspected
    to have multiple good local minima.
}
\end{center}
\end{table}

\section{Two band problem}%
\label{appendix:two_band}

For completeness, we evaluate the \TDC~spread and center 
for the one dimensional two-band problem and remark that 
the \TDC~spread is also continuous with respect to the gauge
\begin{equation}
    U(\theta) = \begin{pmatrix}
        \cos(\theta) & \sin(\theta)\\
        -\sin(\theta) & \cos(\theta)
    \end{pmatrix}.
\end{equation}

\begin{figure}[h]
\begin{center}
    \includegraphics[width=7.8cm]{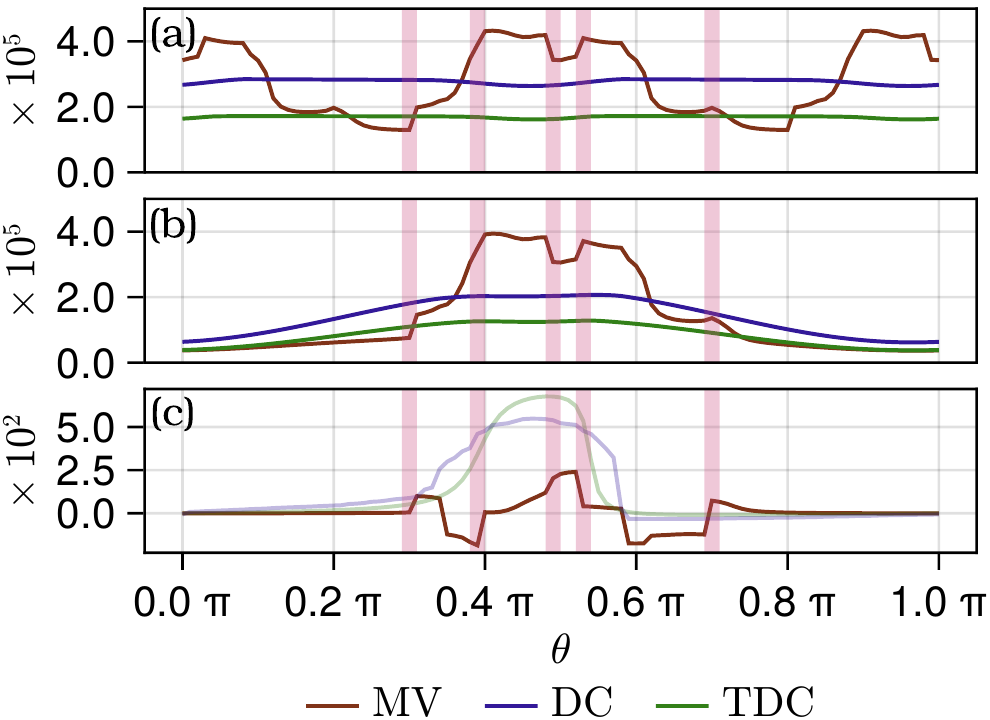}
\end{center}
\caption{A simple two-band problem with eight $k$-points in 1D
    and the \TDC~spread and center included.}
\label{fig:two_band}
\end{figure}

\end{document}